\documentclass[10pt,notitlepage,a4paper,aps,prd,onecolumn,superscriptaddress,nofootinbib,groupedaddress]{revtex4-1}
\usepackage{array}

\usepackage[dvipsnames]{xcolor}

\usepackage[pdfencoding=auto, psdextra]{hyperref}
\usepackage[T1]{fontenc}
\usepackage[utf8]{inputenc}
\usepackage[english]{babel}
\usepackage{hyperref}
\usepackage[mathscr]{euscript}
\usepackage{caption}
\usepackage{subcaption}
\hypersetup{
colorlinks = true,
linkcolor = blue,
filecolor = magenta,
urlcolor = blue,
pdftitle = {TLSs interaction},
pdfpagemode = FullScreen,
citecolor = blue}
\usepackage{physics}

\usepackage{caption}

\usepackage{indentfirst}

\usepackage[pdftex]{graphicx}

\usepackage{amsmath}
\allowdisplaybreaks  
\usepackage{amssymb} 
\usepackage{amsfonts} 
\usepackage{enumerate}

\usepackage{fancyhdr}                     

\usepackage{float}                             

\usepackage{amsmath}         
\usepackage{amsfonts}        
\usepackage{amssymb}         
\usepackage{amsthm}          
\usepackage{empheq}          

\usepackage{color}           
\usepackage{ragged2e}        
\usepackage{titlesec}        

\usepackage{tipa}            

\usepackage[utf8]{inputenc} 
\allowdisplaybreaks          

\usepackage{enumitem}        

\usepackage{tikz}            
\usetikzlibrary{shapes.geometric,arrows.meta,arrows,positioning}  
\usetikzlibrary{tikzmark}   

\titleformat{\paragraph}
  {\normalfont\normalsize\bfseries}{\theparagraph}{1em}{} 
\titlespacing*{\paragraph}
  {0pt}{3.25ex plus 1ex minus .2ex}{1.5ex plus .2ex} 

\begin{document}
\title{Dissipative evolution of a two-level system through a geometry-based classical mapping}

\author{Daniel Martínez-Gil}
\email{daniel.martinez@ua.es}
\affiliation{Fundacion Humanismo y Ciencia, Guzmán el Bueno, 66, 28015 Madrid, Spain.}
\affiliation{Departamento de F\'{\i}sica Aplicada, Universidad de Alicante, Campus de San Vicente del Raspeig, E-03690 Alicante, Spain.}

\author{Pedro Bargueño}
\email{pedro.bargueno@ua.es}
\affiliation{Departamento de F\'{\i}sica Aplicada, Universidad de Alicante, Campus de San Vicente del Raspeig, E-03690 Alicante, Spain.}

\author{Salvador Miret-Artés}
\email{s.miret@iff.csic.es}
\affiliation{Instituto de Física Fundamental, Consejo Superior de Investigaciones Científicas, Serrano 123, 28006, Madrid, Spain}

\begin{abstract}
    In this manuscript, we introduce a geometry-based formalism to obtain a Meyer-Miller-Stock-Thoss mapping in order to study the dynamics of both isolated and interacting
two-level systems. After showing the description of the isolated case using canonically conjugate variables, we implement an interaction model by 
bilinearly coupling the corresponding population differences {\it à la} Caldeira-Leggett, showing that the dynamics behave as a Gross-Pitaevskii-like one.
We also find a transition between oscillatory and tunneling-suppressed dynamics that can be observed by varying the coupling constant.
After extending our model to the {\it system plus environment} case, where the environment is considered as a collection of two-level systems, we show tunneling-suppressed dynamics in the strong coupling limit and the usual damping effect similar to that of a harmonic oscillator bath in the weak coupling one. Finally, we observe that our interacting model turns an isolated symmetric two-level system into an environment-assisted asymmetric one.
\end{abstract}

\maketitle

\section{Introduction}

The exact study of isolated multi-level systems, despite its wide applicability in several subfields of physics and chemistry, is a formidable task. In addition, 
when the environment is considered in order to implement a more realistic dynamics, the model gets even more complicated. In this sense, the following approaches can be made to diminish the complexity of the problem under study: (i) only the two lowest states of the multi-level system are retained (two-level system approximation), (ii) a simple model of the environment and of the system-environment coupling is considered, (iii) a classical description of the dynamics instead of the full quantum one is taken into account. In principle, this classical description might be easily interpreted and, in addition, it has less computational cost.

The interaction between two-level systems (TLSs) and the environment has been widely studied, for instance, in condensed phase dynamics \cite{Kofman, Nitzan_2006}, quantum computing \cite{Chuang_Laflamme_Shor_Zurek_1995, Kendon_2007, Schlosshauer_2019}, spectroscopy in quantum impurity systems \cite{Zhang_Xu_Zheng_Yan_2015,Du_Wang_Xu_Zhang_Yan_2020, Chen_Wang_Xu_Yan_2021}, quantum technologies \cite{QT1, QT2, QT3,QT4,QT5,QT6} or the energy transport in photosyntesis \cite{Duan_Jha_Chen_Tiwari_Cogdell_Ashraf_Prokhorenko_Thorwart_Miller_2022, Lambert_Chen_Cheng_Li_Chen_Nori_2013}. In particular, two types of environments are usually considered: harmonic oscillators and spins (see a comparison between these two types in \cite{Gelman}). 

On one hand, among harmonic oscillator environments, the most popular approach is the Caldeira-Legett model \cite{caldeiraleggett1, Caldeiralegget2, CaldeiraLegget3}, in which the harmonic oscillators are linearly coupled with the system. An example of a harmonic oscillator bath is the spin-boson model, in which the system is considered as a TLS  \cite{RevModPhys.59.1,THORWART2004333, PhysRevA.68.034301, PhysRevB.71.035318}.  In these models, the bath can be reduced to the well-known generalized Langevin equation of motion \cite{weiss}, including both a damping effect and a random external force.

On the other hand, the case of a central TLS in a bath of spins is usually known as the central spin model, which was first introduced by Gaudin in Ref. \cite{Gaudin}.
Although the most general use of these models is to study dissipative processes (see, for example, Refs \cite{PhysRevLett.81.5710,Wangshao, HsiehCao}), there are also specific applications in quantum computing and quantum information, for example with trapped ions \cite{trappedions1, quantumions2} or quantum dots \cite{QuantumDots1, Quantumdots2, Quantumdots3, Quandots4},
the study of these types of models is also of great importance in different physical systems as semiconductors \cite{semiconductors1, semiconductors2} or 
superconductors \cite{superconductor1}.

Regarding the relation between quantum and classical mechanics, the classical limit is usually identified with $\hbar \rightarrow 0$, but the link between Poisson brackets and commutators is difficult to explain in that approximation. Interestingly, Strocchi stated \cite{RevModPhys.38.36}, using a formulation based on complex canonical coordinates, that the Schrödinger equation may be written as Hamilton equations, being the Hamiltonian function the averaged value of the Hamiltonian operator. Subsequently, Meyer and Miller proposed a classical mapping of an F-electronic-state Hamiltonian \cite{Meyermiller}, and Stock and Thoss derived the Meyer-Miller Hamiltonian from Schwinger's theory of angular momentum \cite{StockThoss}. Since then, several authors have proposed other alternative 
quantum-classical mappings. For example, Cotton and Miller \cite{CottonMiller} proposed a spin-mapping (which they claimed to be the ``most natural mapping choice''), which does a nice work for weak couplings \cite{JianLiu}. In subsequent years, Liu showed a unified theoretical framework to construct equivalent representations of the multistate Hamiltonian, presenting six different mapping models \cite{JianLiu}. In other recent works by Runeson and Richardson \cite{Richardson1, Richardson2}, they
 used the Stratonovich-Weyl transform to construct a classical phase space, showing equivalent dynamics to the Meyer-Miller-Stock-Thoss Hamiltonian under a quasiclassical approximation, from the TLS problem to the N-level case. 

In this work, we tackle the problem of considering a TLS interacting with an environment of TLSs using conjugate coordinates variables motivated by a geometric Hopf-based formalism, obtaining a Meyer-Miller-Stock-Thoss (MMST) Hamiltonian. 
Specifically, this work is organized as follows: in Section \ref{section2}, we introduce and motivate a geometric Hopf-based formalism. In Section \ref{section3} we present how an isolated TLS can be expressed in conjugate coordinates variables. To understand our interacting model, we describe a pair of TLSs coupled through their population differences using a Caldeira-Legget-like approach
in Section \ref{section4}. In Section \ref{section5} we study the dynamics of a central TLS interacting bilinearly with an environment of TLSs along the lines presented in previous sections. Finally, Section \ref{section6} provides the conclusions of the present work.

\section{Hopf-based formalism for two level systems}\label{section2}
It is well-known that if $\ket{\psi}$ represents a normalized n-level system, then $\ket{\psi} \in S^{ 2n-1}$, which shows a relation between finite-dimensional Hilbert spaces and odd-dimensional spheres. For instance, if $|\psi\rangle$ describes a TLS (qubit), then $\ket{\psi} \in S^3$. Although this fact seems to contradict the well-known idea of a qubit living on the surface of the Bloch sphere ($S^2$), let us remember that $|\psi '\rangle \sim e^{i \varphi}|\psi\rangle$, where $\sim$ is
an equivalence relation, which reflects the idea that only probabilities (but no amplitudes) matter.

This idea of only considering probabilities but no amplitudes to descend from $S^3$ to $S^2$ can be beautifully described using the first Hopf fibration 
\cite{Hopf1964, Mosseri_2001,URBANTKE2003125,PhysRevA.87.012125}. Basically, the idea is the following: our {\it complete space} ($S^3$) is formed by gluing 
certain {\it fibres} ($S^1$) to our {\it base space} ($S^2$). Note that in this case we are dealing with a {\it non-trivial fibration}, meaning that $S^3 \neq
S^2 \times S^1$ (in this sense, $S^3$ is globally different than $S^2\times S^1$ because there are twistings in the gluing process). An example of a  {\it trivial fibration} is the cylinder, 
$\mathrm{Cyl}=\mathbb{R}\times S^1$, which can be formed gluing $\mathbb{R}$ fibres to the base space, which is $S^1$, without twisting the fibres.

Therefore, the following map from $S^3$ to the Bloch sphere $S^2$, given by $\Pi: S^3 \stackrel{S^1}{\longrightarrow}S^2$, eliminates the global quantum mechanical phase, which is irrelevant to measuring probabilities, as can be easily exemplified in the following example:

If we represent the wave function of a TLS system with a global phase factor as $\ket{\psi'} \sim e^{i\varphi}\ket{\psi} = e^{i\varphi } \alpha \ket{0} + e^{i\varphi} \beta\ket{1}$, the probability to be, for example, in the $\ket{0}$ and $\ket{1}$ states is
\begin{align}
    P_{\ket{0}} & = (e^{i\varphi} \alpha)^*(e^{i\varphi} \alpha) = e^{-i\varphi} \alpha^* e^{i\varphi} \alpha = \alpha^*\alpha = \abs{\alpha}^2,\\
     P_{\ket{1}} & = (e^{i\varphi} \beta)^*(e^{i\varphi} \beta) = e^{-i\varphi} \beta^* e^{i\varphi} \beta = \beta^*\beta = \abs{\beta}^2,
\end{align}
existing invariance under a global phase. In this sense, the $\Pi$ map previously described can be used to provide a geometric way to study the dynamics of TLSs. 
As an aside remark, we would like to point out to the interested reader that there are two other Hopf fibrations, $S^7 \stackrel{S^3}{\longrightarrow}S^4$, and  $S^{15} \stackrel{S^7}{\longrightarrow}S^8$, which can be used in other contexts \cite{Mosseri_2001, S15}.

Let us now detail how these ideas can be formalized in the context of TLSs. In the context of the first Hopf fibration, it can be shown that the result of applying the $\Pi$ map can be expressed in terms of the Pauli matrices as 
\begin{equation}
    \Pi (e^{i\varphi} \ket{\psi} \in S^3) = (\bra{\psi}\hat{\sigma}_x\ket{\psi}, \bra{\psi}\hat{\sigma}_y\ket{\psi}, \bra{\psi}\hat{\sigma}_z\ket{\psi})\in S^2,
\end{equation}
where $\bra{\psi}\hat{\sigma}_x\ket{\psi}^2+\bra{\psi}\hat{\sigma}_y\ket{\psi}^2+\bra{\psi}\hat{\sigma}_z\ket{\psi}^2 = 1$ and 
\begin{align}
    \hat{\sigma}_x & = \begin{pmatrix}
0 & 1\\
1 & 0
\end{pmatrix},
&  \hat{\sigma}_y &=
\begin{pmatrix}
0 & -i\\
i & 0
\end{pmatrix},
& \hat{\sigma}_z &=
\begin{pmatrix}
1 & 0\\
0 & -1
\end{pmatrix}.
\end{align}

Therefore, the set of states $e^{i\varphi} \ket{\psi} \in S^3$ ($\varphi \in [0, 2\pi)$), are mapped to a point on $S^2$ with coordinates $(X, Y, Z)$ \cite{Mosseri_2001}. These coordinates of the Bloch sphere can be expressed as
\begin{align}
    X & = \bra{\psi}\hat{\sigma}_x\ket{\psi} =  2 \mathrm{Re}(\alpha^*\beta),\\
    Y & = \bra{\psi}\hat{\sigma}_y\ket{\psi} =  2 \mathrm{Im}(\alpha^*\beta),\\
    Z & = \bra{\psi}\hat{\sigma}_z\ket{\psi} =  \abs{\alpha}^2-\abs{\beta}^2.
\end{align}
\\
\\
If $\alpha$ and $\beta$ are expressed in polar form ($\alpha = \abs{\alpha}e^{i\phi_\alpha}, \beta = \abs{\beta}e^{i\phi_\beta} \in \mathbb{C}$), we can define 
\begin{equation}
 z \equiv \abs{\alpha}^2-\abs{\beta}^2   
\end{equation}
as the population difference and 
\begin{equation}
  \phi \equiv \phi_\alpha-\phi_\beta  
\end{equation}
as the phase difference.
\\
\\
With these definitions at hand, we can define a pair of canonically conjugate coordinates ($z,\phi$) on $S^2$, expressing the coordinates of the Bloch sphere as 
\begin{align}\label{XYZ}
    X &= \sqrt{1-z^2}\cos \phi, & Y &= \sqrt{1-z^2}\sin \phi, & Z &= z.
\end{align}

Finally, the usual Hamiltonian of a TLS, represented by a combination of Pauli matrices as
\begin{equation}\label{H12}
    \hat{H} = \sum_i \eta_i \hat{\sigma}_i, \eta_i \in \mathbb{R},
\end{equation}
can be Hopf-mapped to its expectation value Hamiltonian function given by
\begin{equation}\label{HamilHopf}
    H_0 = \bra{\psi}\hat{H}\ket{\psi} = -2\eta_1\sqrt{1-z^2}\cos\phi-2\eta_2\sqrt{1-z^2}\sin\phi+2\eta_3 z.
\end{equation}
\\
\\
As a final comment in this section, we would like to remark how our mapping is related to the MMST one. On one hand, Meyer and Miller proposed a mapping of an F-electronic-state Hamiltonian \cite{Meyermiller} by 
\begin{align}\label{MMmapping}
    \hat{H} &= \sum_{m,n = 1}^F H_{nm} \ket{n}\bra{m} &\longrightarrow & & H_{MM} = \bra{\psi}\hat{H}\ket{\psi}
\end{align}

On the other hand, Stock and Thoss \cite{StockThoss} proposed that 
\begin{equation}
    \hat{H} = \sum_{m,n = 1}^F H_{nm} \ket{n}\bra{m} = \sum_{m,n = 1}^F H_{nm} \hat{a}_n^\dagger \hat{a}_m,
\end{equation}
where
\begin{equation}
    \hat{a}_n = \frac{1}{\sqrt{2}} (\hat{X}_n + i\hat{P}_n).
\end{equation}

Therefore, the mapping can be obtained by 
\begin{align}
        \hat{H} &= \hat{H} (\hat{X}, \hat{P}) &\longrightarrow & & H_{ST} = H_{ST}(X,P).
\end{align}

Although the two processes are different, the Hamiltonian function is the same ($H_{MM} = H_{ST}$), being usually known as the Meyer-Miller-Stock-Thoss (MMST) Hamiltonian.

As we are working on the TLS case, our Hamiltonian operator can be expressed as a sum of Pauli matrices (see Eq. \eqref{H12}). Then, our mapping can be expressed as
\begin{align}
  \hat{H} = \sum_i \eta_i \hat{\sigma}_i & &\longrightarrow & & H_{0} = \bra{\psi}\hat{H}\ket{\psi},
\end{align}
which can be easily related to the Meyer-Miller mapping \eqref{MMmapping} in the TLS case. Even more, Cotton and Miller \cite{CottonMiller} have also pointed out that the result of substituting the spin operator $\hat{S}_i = \frac{1}{2}\hat{\sigma_i}$ in their Hamiltonian by the classical angular momentum vector is a MMST map in a specific case. This mapping can be expressed as
\begin{align}
  \hat{H} = \hat{H}(\hat{S}_x, \hat{S}_y, \hat{S}_z) & &\longrightarrow & & H_{CM} = H_{CM}(S_x, S_y, S_z),
\end{align}
where, in action-angle variables and in the MMST-specific case, the classical angular momentum vector is

\begin{equation}
    \begin{pmatrix}
    S_x \\
    S_y \\
    S_z
\end{pmatrix} = 
    \begin{pmatrix}
    \sqrt{1-z^2}\cos \phi \\
    \sqrt{1-z^2}\sin \phi \\
    z
\end{pmatrix}.
\end{equation}

It can also be noted that our mapping fulfills the previous relationship. Therefore, we have obtained a Hamiltonian function \eqref{HamilHopf} given as a sum of coordinates of the Bloch sphere using canonically conjugate variables (see Eq. \eqref{XYZ}). These coordinates were directly obtained by applying the Hopf map $\Pi(e^{i\phi}\ket{\psi})$, thus establishing the relation between the $\Pi$ map and the MMST Hamiltonian.

\section{Description of a TLS using canonically conjugate variables}\label{section3}
In this section, we will represent the Hamiltonian of a TLS by 
\begin{equation}\label{hamilsigmas}
    \hat{H} = \delta \hat{\sigma_x}+\epsilon\hat{\sigma_z},
\end{equation}
which is particularly important, for example, in the context of chiral molecules \cite{Harrisystodolsky}. We denote their eigenfunctions as $\ket{+}$ and $\ket{-}$, which are delocalized states. In order to work with localized states, another algebraic base must be defined (the left and right states of the TLS)
\begin{equation}\label{+-}
    \begin{pmatrix}
\ket{L} \\
\ket{R}
\end{pmatrix}
=
\begin{pmatrix}
\sin\theta & \cos\theta\\
\cos\theta & -\sin\theta
\end{pmatrix}
    \begin{pmatrix}
\ket{+} \\
\ket{-}
\end{pmatrix},
\end{equation}
and $\theta$ is obtained by
\begin{equation}
    \tan2\theta = \frac{2 H_{RL}}{H_{LL}-H_{RR}} = \frac{\delta}{\epsilon},
\end{equation}
where we employ the notation $\bra{X}\hat{H}\ket{Y} \equiv H_{XY}$. Therefore, $\delta$ and $\epsilon$ can be defined in terms of the energy difference between $\ket{+}$ and $\ket{-}$ states, and  between $\ket{L}$ and $\ket{R}$ states as \footnote{$\epsilon$  is the difference between the expected value of the energy of the L and R states, but in the literature is
often referred to as energy difference due to an abuse of language}
\begin{align}
    \delta &= E_+-E_-,& \epsilon &= H_{LL}-H_{RR}.
\end{align}

\begin{figure}[ht]
\centering
\includegraphics[width= 15cm]{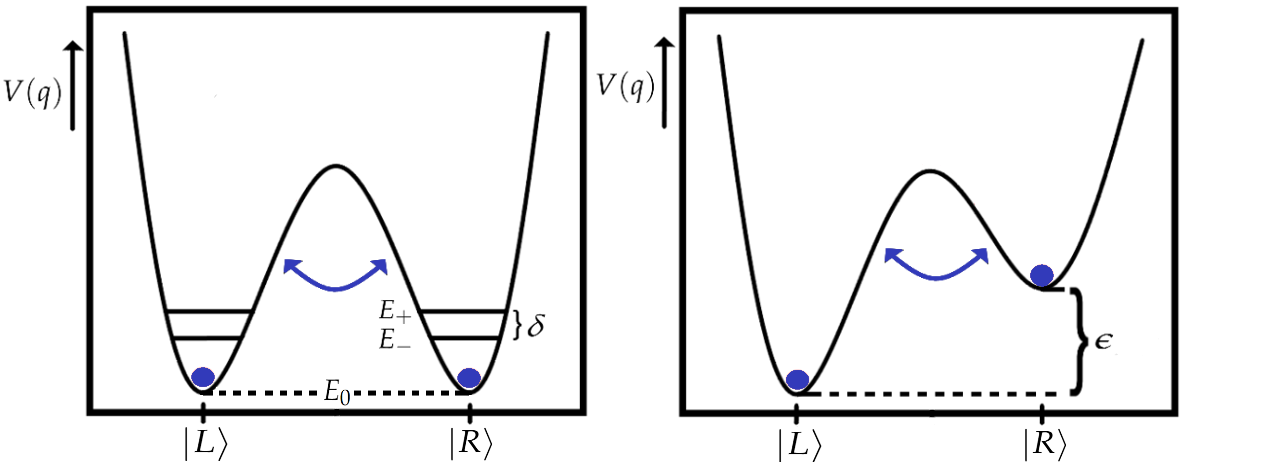}
\caption{\justifying In the left panel, the role of the tunneling effect between the two wells is represented by $\delta$. In the right panel, the asymmetry of the double well potential (as a function of certain generalized coordinate, $q$) is appreciable, which is the consequence of the $\epsilon$ parameter.}
\label{deltaepsilon}
\end{figure}

The $\delta$ parameter is related to the tunneling between the left and right states (is inversely proportional to the tunneling time), and $\epsilon$ is related to the asymmetry in the energy of the left and right states, which can be better understood from Fig. \eqref{deltaepsilon}.

With this in mind, we can express the wave function of the TLS as
\begin{equation}
    \ket{\psi(t)} = a_L(t)\ket{L}+a_R(t)\ket{R},
\end{equation}
obtaining the following time-dependent coupled Schrödinger equations
\begin{align}
    i\hbar \dot{a}_L & = \epsilon a_L +\delta a_R, \\
    i\hbar \dot{a}_R & = \delta a_L -\epsilon a_R.
\end{align}
\\
\\
These equations can be formally mapped to a classical-like system. We can express $a_L, a_R$ in their polar form as $a_{L, R}(t) = \abs{a_{L,R}(t)} e^{i\phi_{L,R}(t)}$, and we can define the population difference and phase difference as was previously done ($z \equiv \abs{a_R(t)}^2-\abs{a_L(t)}^2$ and $\phi(t) \equiv \phi_L(t) - \phi_R(t)$) \cite{PedroCPL2011}.
Using $z$ and $\phi$, we can express the Schrödinger equation by considering them a pair of canonically conjugable variables, which act as a generalized momentum ($z$), and a generalized position ($\phi$). With these changes of variables we rewrite the Schrödinger equation as
\begin{align}
    \dot{z} &= -2\delta\sqrt{1-z^2} \sin \phi \label{z}\\
    \dot{\phi} &= 2\delta \frac{z}{\sqrt{1-z^2}}\cos \phi + 2\epsilon\label{phi}.
\end{align}

\begin{figure}[H]
	\begin{subfigure}[b]{0.49\textwidth}
		\includegraphics[width=\textwidth, height=7cm]{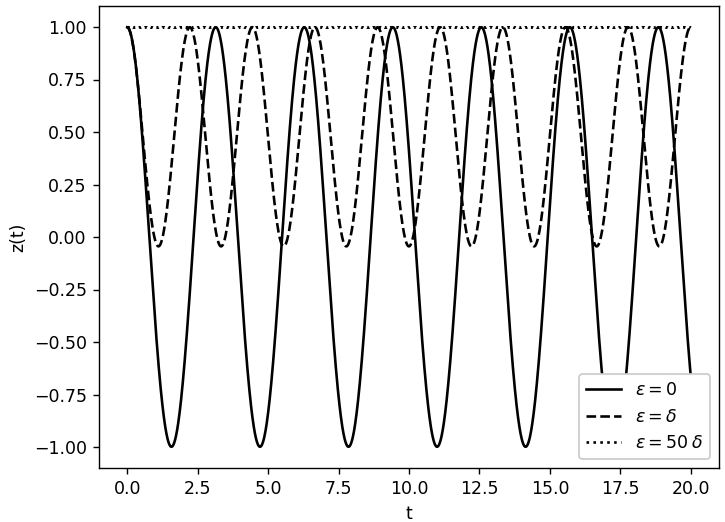}
	\end{subfigure}
	\hfill
	\begin{subfigure}[b]{0.49\textwidth}
		\includegraphics[width=\textwidth, height=7cm]{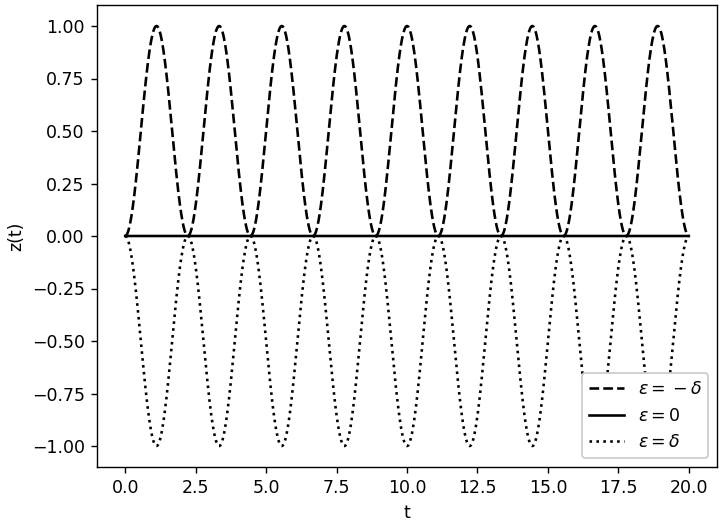}
	\end{subfigure}\caption{In both panels, the population difference between the L and R states is represented for different values of $\epsilon$. In the left 
 and right panels, we begin with $z(0) = 0.999$ and $z(0) = 0$ (racemic mixture), respectively. 
}\label{figura1}
\end{figure}

After integrating Hamilton equations ($\dot{\phi} = \frac{\partial H_0}{\partial z}, \dot{z} = -\frac{\partial H_0}{\partial \phi}$), we obtain
\begin{equation}\label{hamil0}
    H_0 = -2\delta \sqrt{1-z^2} \cos \Phi + 2\epsilon z.
\end{equation}
As an important remark, we note that the previous Hamiltonian is the Meyer-Miller-Stock-Thoss Hamiltonian \eqref{HamilHopf} \cite{Meyermiller, StockThoss} considering $\eta_1 = \delta, \eta_2 = 0, \eta_3 = \epsilon$. Therefore, it can be shown that the Hamiltonian operator \eqref{hamilsigmas} and the Hamiltonian function \eqref{hamil0} are related by doing $H_0 = \bra{\psi}\hat{H}\ket{\psi}$, as was mentioned in Eq. \eqref{HamilHopf}.
\\
\\
At this point, it is interesting to show how Eqs. \eqref{z} and \eqref{phi} exactly reproduce the dynamics of a TLS (Fig. \eqref{figura1}). In the left figure, we can observe how the TLS tends to stabilize in its R state for large values of $\epsilon$, suppressing the tunneling effect. In the right figure, there is no population difference for $\epsilon = 0$. If $\epsilon \neq 0$, depending on its sign, the TLS tends to be located in the L or R state.

Once the isolated model has been reviewed using the aforementioned canonically conjugate variables, let us consider how to introduce interations in the following section.

\section{Interacting two-level systems}\label{section4}

In order to understand our interaction model, let us establish a similarity with the harmonic oscillator bath case. A general mass point $M$ described in phase-space coordinates $X$ (position) and $P$ (momentum) and moving in a potential $V(X))$, interacting with a bath of harmonic oscillators, can be described by the Caldeira-Legget Hamiltonian \cite{caldeiraleggett1, Caldeiralegget2, CaldeiraLegget3}
\begin{equation}
    H = \frac{P^2}{2 M} +V(X) +\sum_i \frac{1}{2}(\frac{p_i^2}{m_i}+ m_iw_i^2 x_i^2) + H_I.
\end{equation}

Avoiding a possible counterterm for the renormalization of the potential \cite{weiss}, and considering bilinear couplings, the interaction Hamiltonian has four possibilities
\begin{align}\label{hamilinteraction}
    H_I &= X\sum_i \Omega_i x_i, &
     H_I &= P\sum_i \Omega_i p_i, &
     H_I &= X\sum_i \Omega_i p_i, &
     H_I &= P\sum_i \Omega_i x_i, 
\end{align}
which are the position-position, momentum-momentum (which is usually known as anomalous coupling \cite{Pollak, Cuccoli, Maile, Pollak2, Ferialdi}), and the mixing couplings ($\Omega_i$ are the corresponding coupling constants).

Inspired by the harmonic oscillator bath case, we introduce our interacting model between two TLSs as follows. The total Hamiltonian will be represented by
\begin{equation}
    H_T = \underbrace{-2\delta_1 \sqrt{1-Z^2} \cos \Phi + 2\epsilon_1 Z}_{H_1}+\underbrace{-2\delta_2 \sqrt{1-z^2} \cos \phi + 2\epsilon_2 z}_{H_2} + H_I,
\end{equation}
where $H_1$ and $H_2$ are similar to the Hamiltonian function \eqref{hamil0}, representing each of them an isolated TLS, and the interaction term is given by $H_I$. As in the harmonic bath case, given by Eqs. \eqref{hamilinteraction}, we propose the following interaction terms:
\begin{align}
    H_I &= \Lambda_{PP} \cos\Phi \cos\phi, &
    H_I &= \Lambda_{MM} Z z, &
    H_I &= \Lambda_{MP} z \cos \Phi, &
    H_I &= \Lambda_{PM} Z \cos \phi, 
\end{align}
which acts as the position-position, momentum-momentum, and mixing couplings and the $\Lambda$s are the corresponding coupling constant. Let us note 
that our position coordinate $\phi$ is under a cosine. As was argued in \cite{Coseno} and \cite{weiss} in the context of Josephson systems, $\phi$ is a phase variable and thus it is expected to be under a trigonometric function.

Considering all the possible interacting Hamiltonians, and applying Hamilton equations for each two-level system ($\dot{\Phi} = \frac{\partial H_T}{\partial Z}, \dot{Z} = -\frac{\partial H_T}{\partial \Phi}, \dot{\phi} = \frac{\partial H_T}{\partial z}, \dot{z} = -\frac{\partial H_T}{\partial \phi}$), we obtain the following differential equations
\begin{eqnarray}
    \dot{Z} =& -\delta_1\sqrt{1-Z^2}\sin \Phi + \Lambda_{PP} \sin \Phi \cos \phi + \Lambda_{MP} z \sin \Phi, \label{Zcontodos1} \\
    \dot{\Phi} =& 2\epsilon_1 + \delta_1 \frac{Z}{\sqrt{1-Z^2}}\cos \Phi +  \Lambda_{MM} z + \Lambda_{PM} \cos \phi,\\
    \dot{z} = & -\delta_2 \sqrt{1-z^2}\sin \phi + \Lambda_{PP} \cos \Phi  \sin \phi + \Lambda_{PM} Z \sin \phi,\label{zcontodos1}\\
    \dot{\phi} =& 2\epsilon_2+ 2\delta_2 \frac{z}{\sqrt{1-z^2}} \cos \phi + \Lambda_{MM} Z  + \Lambda_{MP} \cos \Phi.
\end{eqnarray}

At this point, some comments are in order. Note that some of these interacting terms are not allowed in the following sense. The parameters $Z$ and $z$ must be bounded between $+1$ and $-1$, and all the interaction terms in the Eqs. \eqref{Zcontodos1} and \eqref{zcontodos1} break the boundedness of these variables (this can be shown for example by considering all coupling strengths higher enough, $\Lambda \gg 1$, and $\Phi(0) = \phi(0) = \pi/4$). Therefore, the only interaction term that preserves the boundedness of the population differences is the momentum-momentum coupling. Then the differential equations we are going to consider are 

\begin{eqnarray}
    \dot{Z} =& -\delta_1\sqrt{1-Z^2}\sin \Phi , \label{Zcontodos} \label{1}\\
    \dot{\Phi} =& 2\epsilon_1 + \delta_1 \frac{Z}{\sqrt{1-Z^2}}\cos \Phi +  \Lambda z \label{2}\\
    \dot{z} = & -\delta_2 \sqrt{1-z^2}\sin \phi,\label{zcontodos}\label{3}\\
    \dot{\phi} =& 2\epsilon_2+ 2\delta_2 \frac{z}{\sqrt{1-z^2}} \cos \phi + \Lambda Z, \label{4}
\end{eqnarray}
where we have denoted $\Lambda = \Lambda_{MM}$. 

Another interesting point to consider is the similarity between our momentum-momentum coupling and the usual spin-spin coupling \cite{Schlosshauer_2019}.
Specifically, we can write
\begin{equation} \label{interactionhamil}
    H_I = \Lambda Z z = \Lambda \bra{\psi}\hat{\sigma}_{z}\ket{\psi}\bra{\phi}\hat{\sigma}_{z}\ket{\phi} = \langle\hat{\sigma}_{z}\rangle_\psi\langle\hat{\sigma}_{z}\rangle_\phi,
\end{equation}
where $\psi$ and $\phi$ are the wave function of each TLS. Therefore, the total Hamiltonian can be expressed as 
\begin{equation*}
    H_T = -2\delta_1 \sqrt{1-Z^2} \cos \Phi + 2\epsilon_1 Z+-2\delta_2 \sqrt{1-z^2} \cos \phi + 2\epsilon_2 z + \Lambda z Z =
\end{equation*}
\begin{equation}\label{HT}
    = \epsilon_1 \langle \hat{\sigma}_{z}\rangle_{\psi}+\delta_1 \langle\hat{\sigma}_{x}\rangle_{\psi} + \epsilon_2\langle\hat{\sigma}_{z}\rangle_{\phi} +\delta_2\langle\hat{\sigma}_{x}\rangle_{\phi} + \Lambda \langle\hat{\sigma}_{z}\rangle_{\psi}\langle\hat{\sigma}_{z}\rangle_{\phi},
\end{equation}
which is, somehow, similar to the spin-spin interaction, this time introducing state-averaging.

As a final comment, we want to remark that our interacting Hamiltonian \eqref{interactionhamil} couples bilinearly the population differences of each TLS, which act as the corresponding momentum coordinates. In this sense, we are implementing an anomalous coupling between the two-level systems.

\subsection{Results}
In this section, we will show how our interacting model behaves under certain conditions, which could be understood, for example, as a simple model of two chiral molecules interacting through their population difference. 

\subsubsection{The role of the coupling constant}

\begin{figure}[ht]
\centering
\includegraphics[width= 11cm]{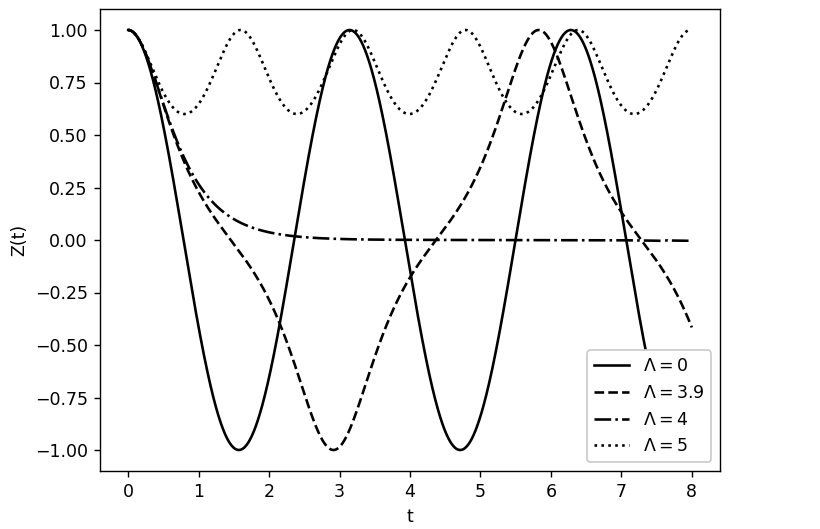}
\caption{\justifying In this picture we show the dynamics of two interacting  TLS using Eqs. \eqref{Zcontodos}-\eqref{4}. We use $\epsilon_1 = \epsilon_2 = 0$, $\delta_1 = \delta_2 = 1$, $Z(0) = z(0) = 0.99999$, and $\Phi(0) = \phi(0) = 0$.}
\label{2TLS}
\end{figure}

First of all,  in Fig. \eqref{2TLS} we expose the role of the $\Lambda$ parameter.
As we can see, the oscillations become anharmonic and of lower frequency when we augment the value of $\Lambda$, until $\Lambda$ reaches its critical value (in this case, $\Lambda_c = 4$) and the oscillations disappear. If $\Lambda > \Lambda_c$, the blocking of the tunneling effect can be observed, this effect being closely related to the ``self-trapping effect''     of two Bose-Einstein condensates \cite{pendulo1,Pendulo2,Pendulo3}, where the coherent atomic tunneling between two zero-temperature Bose-Einstein condensates (BEC) confined in a double-well magnetic trap can be studied using two Gross-Pitaevskii equations for the self-interacting
BEC amplitudes. In these works, the authors also used the population difference and phase difference as variables, obtaining
\begin{align}
\dot{z} =& -\sqrt{1-z^2}\sin \phi \\
\dot{\phi} =& \Delta E+ \Lambda z  + \frac{z}{\sqrt{1-z^2}} \cos \phi.
\end{align}

Comparing Fig. \eqref{2TLS} with their results, we can conclude that when $\epsilon_1 = \epsilon_2 = 0$ and the population difference and phase difference at the initial time of each TLS are the same, our system behaves similarly to a single non-rigid pendulum \cite{pendulo1,Pendulo2,Pendulo3}. In this case, we can obtain an analytical expression for the critical value of the coupling strength, as we will see.

The Hamiltonian \eqref{hamilinteraction} with $\epsilon_1 = \epsilon_2 = 0$ can be expressed as
\begin{equation}\label{Hamil35}
    H_T = -2\delta_1 \sqrt{1-Z^2} \cos \Phi - 2\delta_2 \sqrt{1-z^2} \cos \phi + \Lambda Z z.
\end{equation}

If we impose that there is time $t_1$ for which $Z(t_1) = z(t_1) = 0$,  then
\begin{equation}\label{hamiltonianot1}
    H(t_1) = - 2 \delta_1 \cos \Phi(t_1) - 2 \delta_2 \cos \phi (t_1).
\end{equation}

Taking into account the conservation of the total Hamiltonian \eqref{hamiltonianot1}, we obtain $\abs{ H(t_1)} < 2\delta_1+2\delta_2$. Therefore, $\abs{H_T} = 2\delta_1+2\delta_2$ is the limit for which $Z(t), z(t)$ will never reach the zero value.
Substituting $\abs{H_T} = 2\delta_1+2\delta_2$ in Eq. \eqref{Hamil35}, we obtain two critical values for $\Lambda$:
\begin{equation}
    \Lambda_c = -\frac{\pm (2\delta_1+2\delta_2) + 2\delta_1\sqrt{1-Z(0)^2}\cos \Phi(0)+2 \delta_2 \sqrt{1-z(0)^2}\cos \phi(0)}{Z(0)z(0)},
\end{equation}
where we express this equation in terms of the initial values of the parameters.

As it can be shown in Fig. \eqref{2TLS}, the dashed-dotted line acts as a transition between oscillatory and tunneling-block dynamics (in this case, the critical value of the coupling strength is $\Lambda_c = 4$). We want to remark that although our results are quite similar to the BEC confined in a double-well magnetic trap, our system is totally different since we consider four differential equations, and our Hamiltonian can not be represented by a single non-rigid pendulum.

It is also interesting to note that the Gross-Pitaevskii equation has also been used to study interacting chiral molecules (a particular case of TLS) in a mean-field theory \cite{vardi}, obtaining a similar result to Fig. \eqref{2TLS}. In this sense, our model could be understood as a Gross-Pitaevskii-like equation where the 
$ \Lambda $-term represents nonlinearities.

\begin{figure}[H]
	\begin{subfigure}[b]{0.49\textwidth}
		\includegraphics[width=\textwidth, height=7cm]{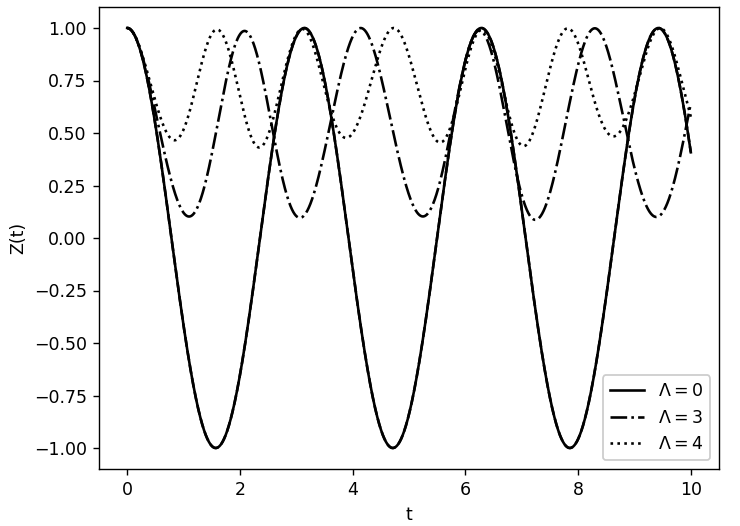}
	
	\end{subfigure}
	\hfill
	\begin{subfigure}[b]{0.49\textwidth}
		\includegraphics[width=\textwidth, height=7cm]{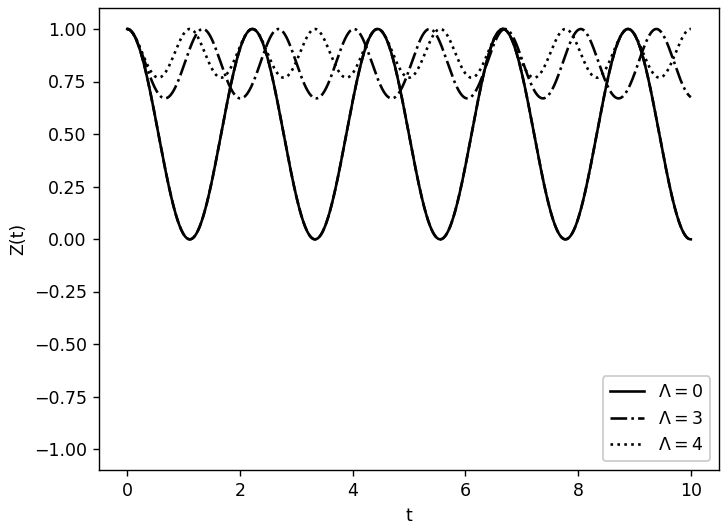}

	\end{subfigure}\caption{This figure is divided into two parts, both of them considering $Z(0) = z(0) = 0.99999$, $\Phi(0) = \phi(0) = 0$ and $\delta_1 = \delta_2 = 1$. In the left one, $\epsilon_1 = 0, \epsilon_2 = 1$. In the right one, $\epsilon_1 = \epsilon_2 = 1$. Compare this results with Fig. \eqref{2TLS}. 
}\label{dosmolconepsilon}
\end{figure}

Interestingly, there is no any {\it sharp} transition-like behaviour similar to that observed in Fig. \eqref{2TLS} when $\epsilon \ne 0$, as 
Fig. \eqref{dosmolconepsilon} shows.

\subsubsection{The role of initial conditions}
\begin{figure}[H]
	\begin{subfigure}[b]{0.49\textwidth}
		\includegraphics[width=\textwidth, height=7cm]{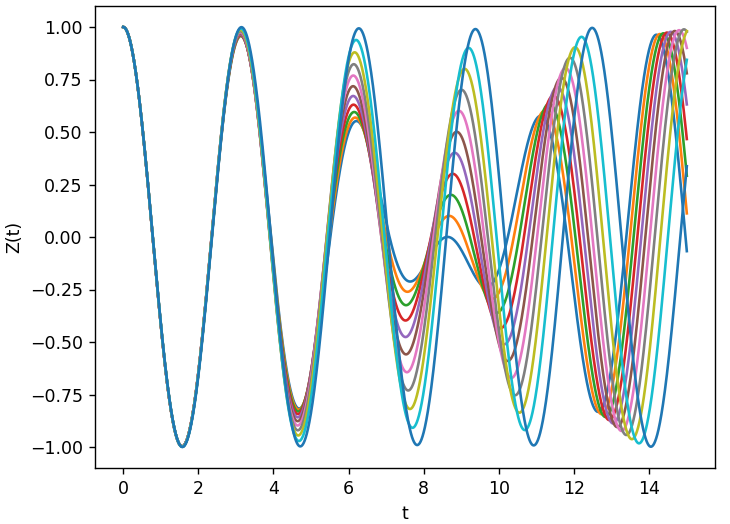}
	\end{subfigure}
	\hfill
	\begin{subfigure}[b]{0.49\textwidth}
		\includegraphics[width=\textwidth, height=7cm]{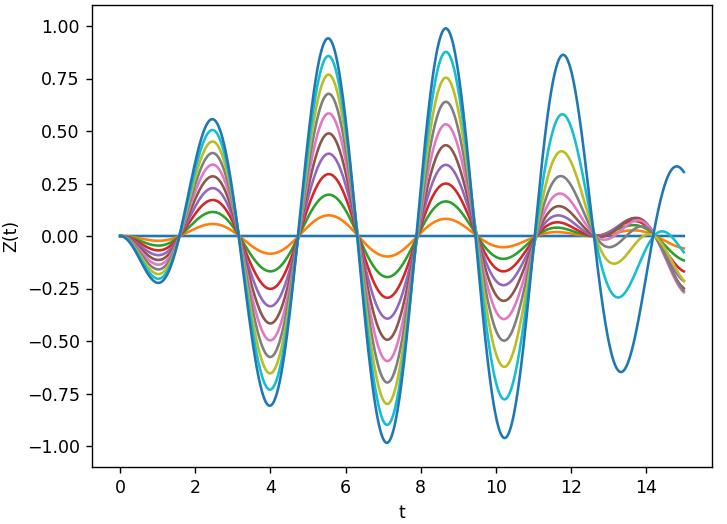}
	\end{subfigure}\caption{
 In both panels, we can see $Z(t)$ varying $z(0)$ between $0$ and $0.99999$, considering $\Lambda = 0.5$. The results with large amplitudes correspond to $z(0) = 0.999999$ and those with small amplitudes correspond to $z(0) = 0$. We also take  $\Phi(0) = \phi(0) = 0$, $\delta_1 = \delta_2 = 1$ and $\epsilon_1 = \epsilon_2 = 0$. We have chosen $Z(0) = 0.999999$ and $Z(0) = 0$ in the left and right panels, respectively.
}\label{variandoz01}
\end{figure}

\begin{figure}[H]
	\begin{subfigure}[b]{0.49\textwidth}
		\includegraphics[width=\textwidth, height=7cm]{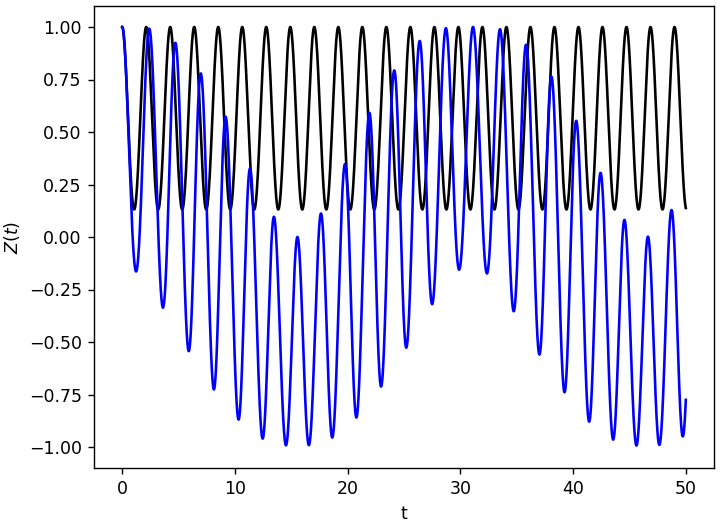}
	\end{subfigure}
	\hfill
	\begin{subfigure}[b]{0.49\textwidth}
		\includegraphics[width=\textwidth, height=7cm]{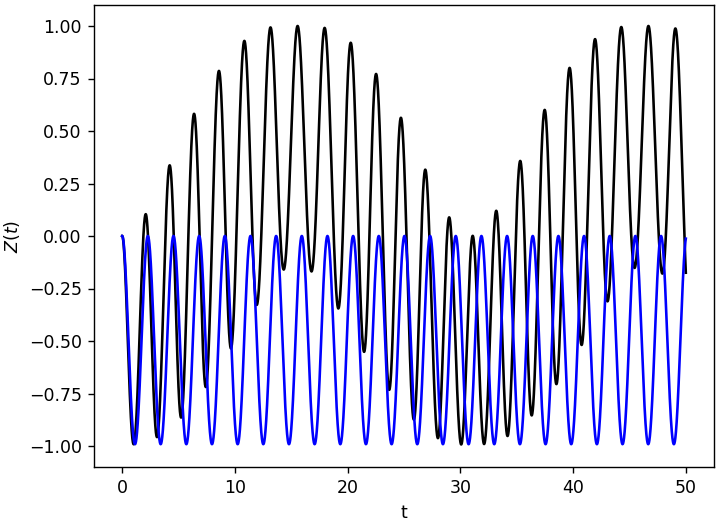}
	\end{subfigure}\caption{We plot $Z(t)$ changing the value of $z(0)$, taking $\Lambda = 0.5$. Black lines correspond to $z(0) = 0.99999$ and blue lines to $z(0) = 0$.  We also consider $\Phi(0) = \phi(0) = 0$, $\delta_1 = \delta_2 = 1$ and $\epsilon_1 = \epsilon_2 = 1$. $Z(0) = 0.99999$ and $Z(0) = 0$ in the left and right panels, respectively. 
}\label{variandoz02}
\end{figure}

Once the role of $\Lambda$ has been clarified, let us vary the initial populations (changing the initial phases do not result in any appreciable results). We can see 
these results in Figs. \eqref{variandoz01} and \eqref{variandoz02}. In the first figure, although a large amplitude is observed when $z(0) \approx 1$, the 
time-averaged value of the population difference is zero for all cases, therefore not having relevant observable conclusions. In Fig. \eqref{variandoz02}, 
once a non-vanishing $\epsilon$ is added,  we can observe (left picture) that $\langle Z(t)\rangle_t \neq 0$ when $z(0) \approx 1$ and $\langle Z(t)\rangle _t = 0$ when $z(0) = 0$. On the contrary, in the right picture, $\langle Z(t)\rangle _t \neq 0$ when $z(0) = 0$ and $\langle Z(t)\rangle _t = 0$ when $z(0) \approx 0$. Therefore, in this case, the initial value of the population difference of the second TLS could have observable effects in the first TLS. Finally, we have to mention that we do not consider negative values neither for $\epsilon$ nor for $\Lambda$ because the conclusions are quite similar to the positive cases.

\section{Environment as a collection of TLSs}\label{section5}
In the previous section, we have presented a model of two interacting TLSs. In this section, we consider the effects of a large number of TLSs by using
 a system + environment formalism in the following form:
\begin{equation}
    H_T = H_S+H_{E}+H_{I},
\end{equation}
where $H_S, H_E, H_I$ refer to the system, environment, and interaction respectively.

\begin{figure}[ht]
\centering
\includegraphics[width= 8cm]{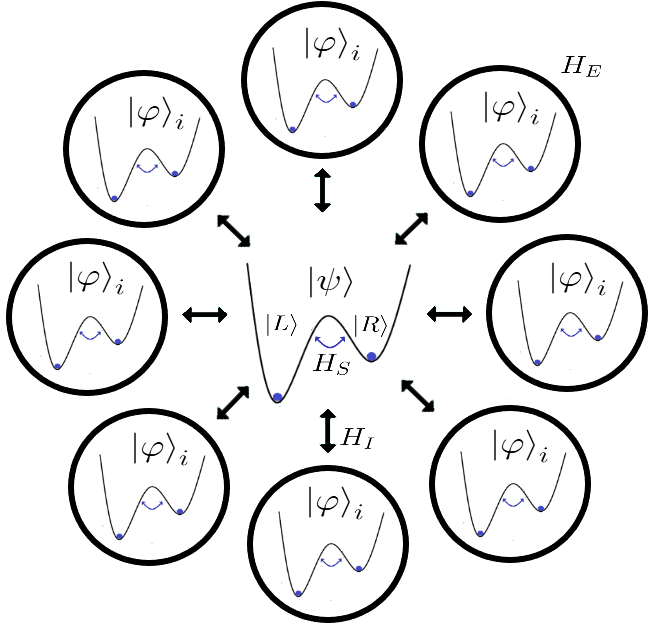}
\caption{\justifying The system + environment model we are considering.}
\label{esquema}
\end{figure}

This formalism has been widely studied in the context of open TLSs by considering the environment as a bath of harmonic oscillators \cite{PedroCPL2011,HarrisSilbey1984, PedroChirality2013, PedroChirality2014, Brito_2008, PhysRevB.72.195410, PhysRevA.94.032127, CHANG1993483} or as a spin bath \cite{PhysRevLett.81.5710,Wangshao, HsiehCao, Golosov,DAVIDMERMIN1991561, PhysRevA.106.032435, PhysRevB.60.972, PhysRevA.72.052113, PhysRevB.101.184307, PhysRevA.71.052321, NVProkofev_2000, Gelman, newspinbath1, newspinbath2, newspinbath3, newspinbath4, newspinbath5, newspinbath6}. In our model, we will consider the environment as a collection of TLSs, which 
interact with the central TLS through a Caldeira-Legget-like formalism \cite{caldeiraleggett1, Caldeiralegget2, CaldeiraLegget3}, neglecting the interaction between the TLSs of the environment (see Fig. \eqref{esquema}).

Therefore, the total Hamiltonian in this system + environment formalism can be written as
\begin{equation}
    H = \underbrace{-2\delta \sqrt{1-Z^2} \cos \Phi + 2\epsilon Z}_{H_S} + \underbrace{\sum_i \left[-2\delta_i\sqrt{1-z_i^2} \cos \phi_i + 2\epsilon_i z_i\right]}_{H_E}+\underbrace{Z\sum_i \Lambda_i z_i}_{H_I},
\end{equation}
where the index $i$ goes from 1 to $N$, being $N$ the total number of TLS in the environment, and $\Lambda_i$ represents the coupling strenght of the central TLS with the $i$th TLS of the environment. Therefore, applying Hamilton equations, we obtain 2$N$+2 differential equations (2 for the system, and 2$N$ for the environment). We can express these equations as

\begin{eqnarray}
    \dot{Z} =& -2\delta\sqrt{1-Z^2}\sin \Phi, \label{Zsystem}\\
    \dot{\Phi} =& 2\epsilon+2\delta \frac{Z}{\sqrt{1-Z^2}}\cos \Phi + \sum_i \Lambda_i z_i,\\
    \dot{z}_i = & -2\delta_i \sqrt{1-z_i^2}\sin \phi_i,\\
    \dot{\phi}_i =& 2\epsilon_i+ 2\delta_i \frac{z_i}{\sqrt{1-z_i^2}} \cos \phi_i + Z\sum_i \Lambda_i.\label{phienvironment}
\end{eqnarray}

Note that our results will not converge for an arbitrary large number of environmental TLSs, $N$, in order to consider them a {\it bath}, unless we introduce
certain dependence of $\Lambda$ on $N$ (for instance, $\Lambda \sim \frac{1}{N}$). Therefore, here we will focus on a particular value of $N$ (say $N=12$),
leaving the continuum case for future work.

\begin{figure}[ht]
\centering
\includegraphics[width= 8cm]{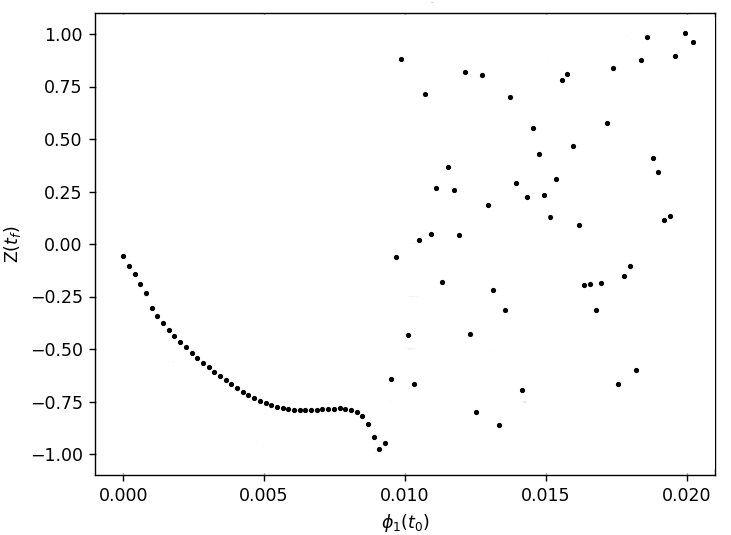}
\caption{\justifying In this figure, we can observe that the system has a non-chaotic behavior when $\phi_1(t_0) < 0.008$. On the contrary, when $\phi_1(t_0) > 0.008$, a chaotic behavior is appreciable, being extrapolable to the $N>2$ case. Here we plot the population difference of the system at a final time ($t_f = 20$) changing the initial phase difference of a TLS of the environment. In this case, $\Lambda = 1$, $N = 2$. }
\label{chaotic}
\end{figure}

As an aside note, observe that the system described by Eqs. \eqref{Zsystem}-\eqref{phienvironment} develops chaotic features for $N > 1$. 
For example, Fig. \eqref{chaotic} shows this chaotic behaviour for $N=2$, which is reminiscent of that of a double pendulum.

Due to this chaotic dynamics, we will employ certain averaging of the population differences to obtain conclusions, as it will explained in the following section. 

\subsection{Results}
Once being clear the model we are using, we can start showing some numerical results. 

\subsubsection{The role of the coupling constant}

\begin{figure}[H]
	\begin{subfigure}[b]{0.49\textwidth}	\includegraphics[width=\textwidth]{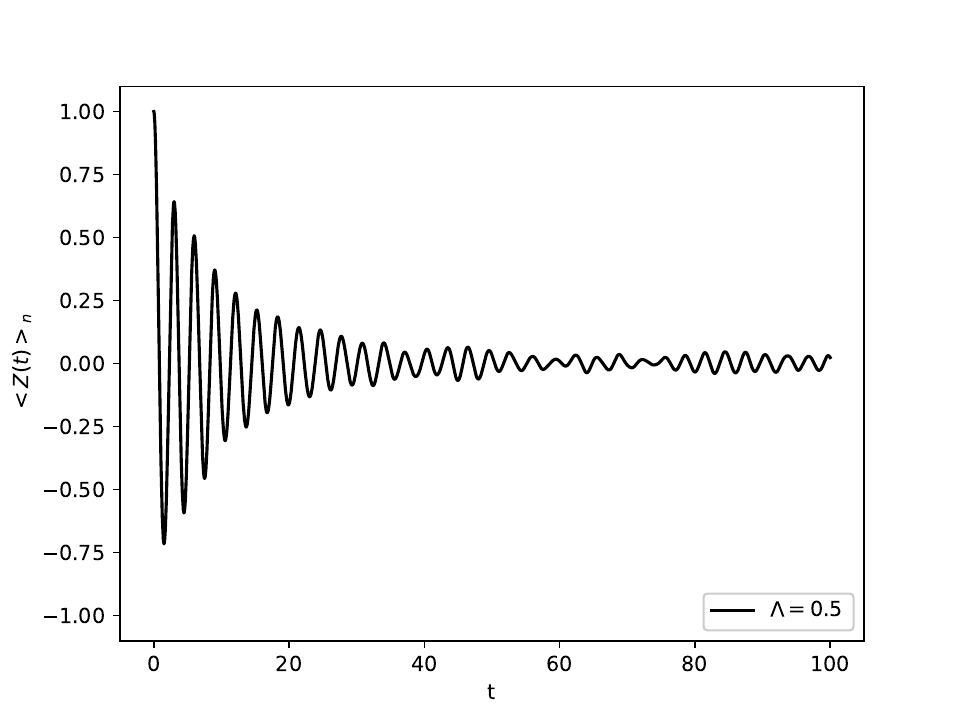}
	\end{subfigure}
	\hfill
	\begin{subfigure}[b]{0.49\textwidth}		\includegraphics[width=\textwidth]{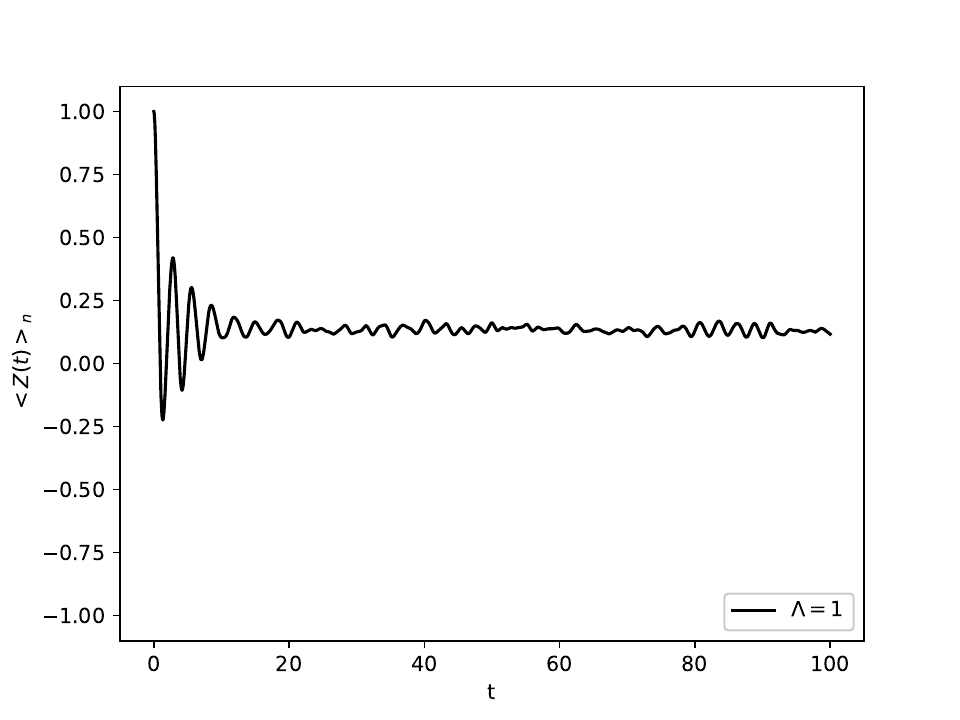}
\end{subfigure}
	\hfill
	\begin{subfigure}[b]{0.49\textwidth}		\includegraphics[width=\textwidth]{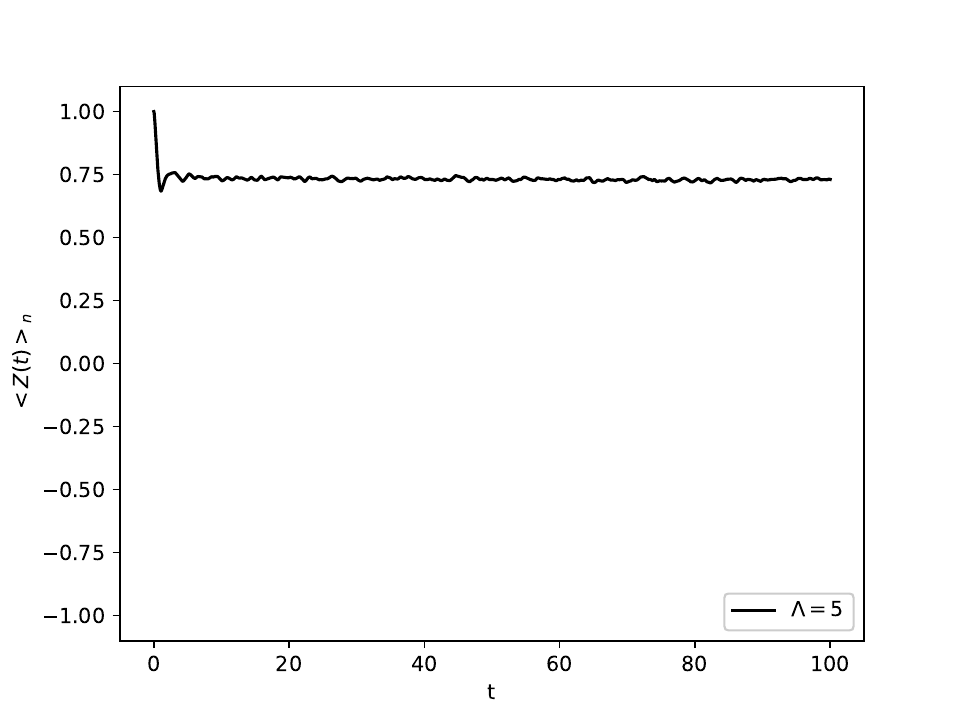}
\end{subfigure}
	\hfill
	\begin{subfigure}[b]{0.49\textwidth}		\includegraphics[width=\textwidth]{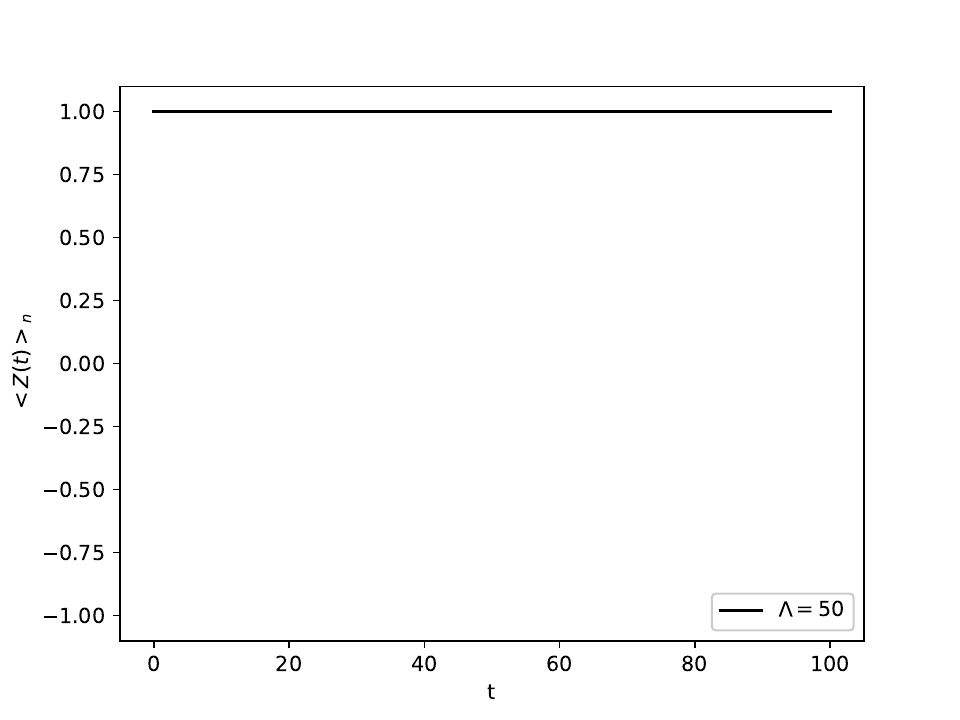}
\end{subfigure}    \caption{We plot the averaged value of the population difference of the system in the number of realizations, $n$, varying the coupling strength $\Lambda$. We have chosen $n = 2000$, $\epsilon = \epsilon_i = 0$, $\delta = \delta_i = 1$, and $z_i(0)$ and $\phi_i(0)$ as random parameters between $(-1,1)$ and $\left[0, 2\pi\right)$ respectively.}
    \label{damping}
\end{figure}

As we have done in the previous section, we will begin by showing how the $\Lambda$ parameter affects the system. We have to remark that in the rest of the manuscript, we will consider $\Lambda_i = \Lambda$ for any TLS of the environment. As we can see in Fig. \eqref{damping}, we are showing $\langle Z(t)\rangle _n$, which refers to the main value of the population difference of the system in the number of realizations, $n$. Although the system is now chaotic, we can obtain conclusions by performing this average. Returning to Fig. \eqref{damping}, tunneling suppression appears when we increase the value of $\Lambda$. In the strong coupling limit, for example, if $\Lambda = 50$, we can see that the system is localized in its R state. With respect to the weak coupling limit ($\Lambda = 0.5$ in Fig. \eqref{damping}), it was shown in \cite{decoherence} that the dynamics of a system interacting with either spin or harmonic oscillators are equivalent \cite{CaldeiraLegget3, FEYNMAN1963118} (note that the first mapping from spin to harmonic oscillator environment was studied in \cite{PhysRevB.48.13974}). Therefore, the Born-Markov master equation for the spin environment can be expressed as the Born-Markov master equation for the oscillator environment evaluated with a modified spectral density.
Although our environment is not a bath {\it per se}, a characteristic damping effect (see for example \cite{PedroCPL2011}) similar to that obtained in the harmonic oscillator bath case is obtained.  

\begin{figure}[H]
	\begin{subfigure}[b]{0.49\textwidth}
	\includegraphics[width=\textwidth]{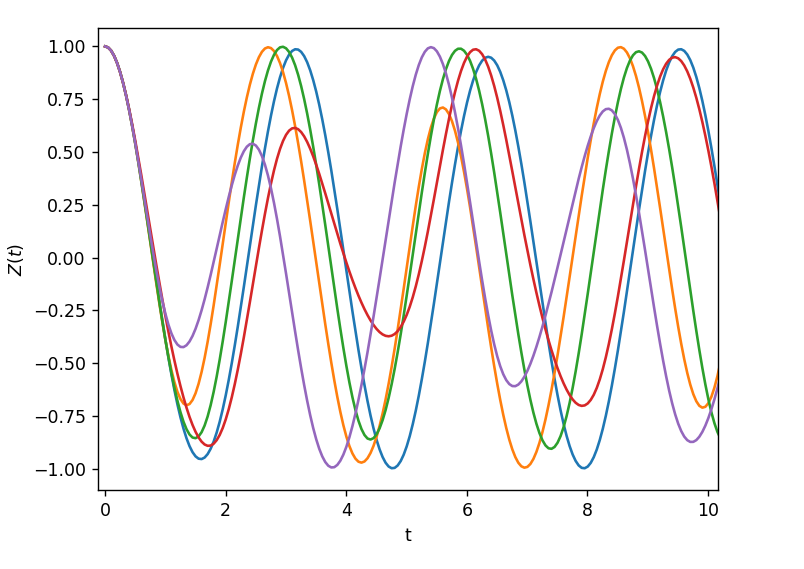}
	\end{subfigure}
	\hfill
	\begin{subfigure}[b]{0.49\textwidth}
    \includegraphics[width=\textwidth]{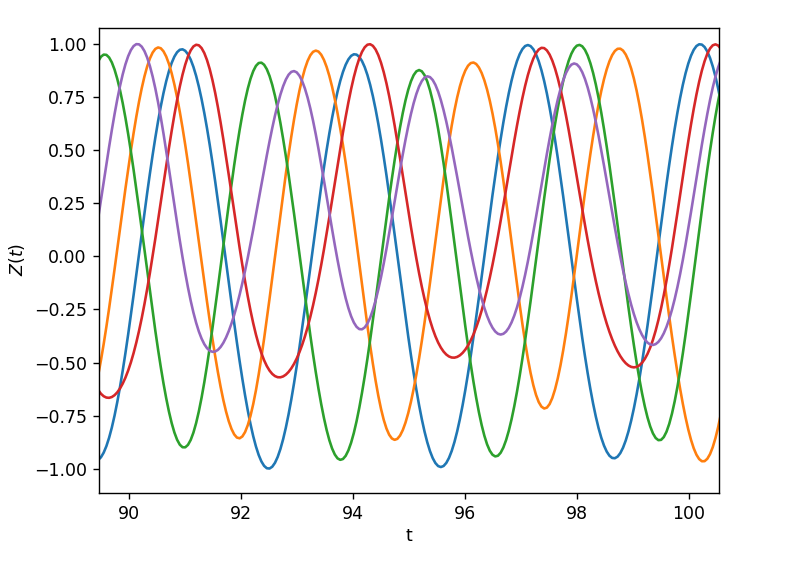}
	\end{subfigure}\caption{5 different simulations considering  $\Lambda = 0.5$ and $z_i(0), \phi_i(0)$ as random parameters. We also consider $Z(0) = 0.9999, \Phi(0) = 0$, $\epsilon = \epsilon_i = 0$ and $\delta = \delta_i = 1$. The initial and final times are shown in the left and right panels, respectively.
}\label{chaotic2}
\end{figure}

In order to show the chaotic behavior of the system, in Fig. \eqref{chaotic2} we plot $n = 5$ realizations considering $\Lambda = 0.5$. As we can observe, while at initial times the different realizations behave similarly, this does not happen at final times due to the chaos of the system. This figure can be used as an illustrative example to understand the dynamics of the top left pannel of Fig. \eqref{damping}.

Another important point to mention refers to the maximum standard deviation of each result. The maximum standard deviation in the $\Lambda = 0.5$ case is $\sigma_{max} = 0.66$, having a high dispersion measure due to the chaotic behavior of the system. It is also appreciable that the maximum standard deviation decreases when we increase the value of $\Lambda$, making the results in the strong coupling limit more reliable ($\sigma_{max} = 0.04$ when $\Lambda = 50$).
\vspace{-1cm}

\subsubsection{The role of $\epsilon$}
\begin{figure}[ht]
\centering
\includegraphics[width= 10cm]{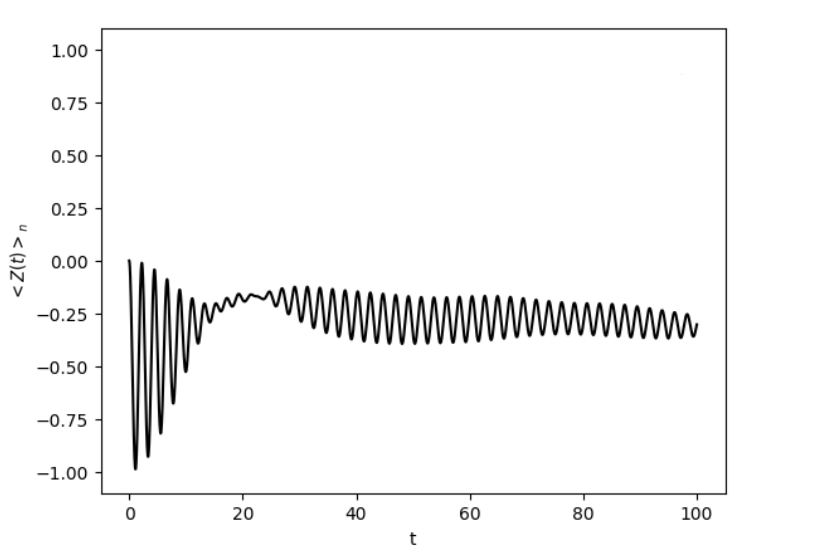}
\caption{\justifying Averaged value of the population difference of the system for a given number of realizations, $n$. In this case, we consider $n = 2000$, $\Lambda = 0.01$, $\epsilon = \epsilon_i = 1$, $\delta = \delta_i = 1$. We also consider  $z_i(0)$ and $\phi_i(0)$ as random parameters as in the previous figure.}
\label{figura10}
\end{figure}

In the previous subsection, we have studied the role of the $\Lambda$ parameter considering $\epsilon= \epsilon_i=0$. That is, no asymmetry neither for the central
nor for the environmental TLSs were considered. After including the aforementioned asymmetries, we observe (see Fig. \eqref{figura10}): (i) a damping effect as a consequence of the couplings and (ii) a non-zero averaged population difference.

\begin{figure}[H]
\centering
\includegraphics[width= 10cm]{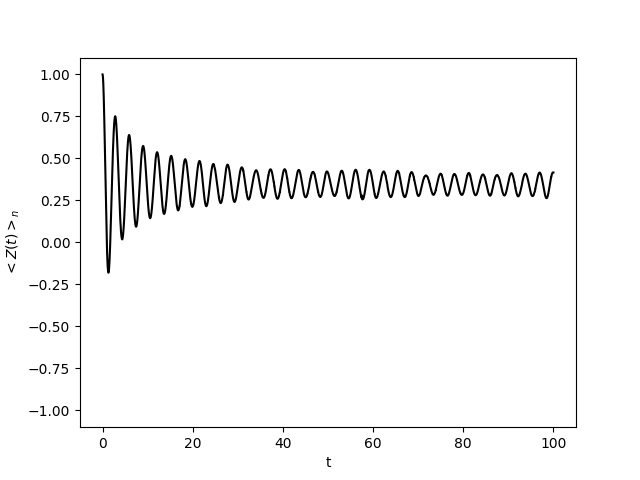}
\caption{\justifying Averaged value of the population difference of the system for a given number of realizations, $n$. In this case, we consider  $n = 2000$, $\Lambda = 1$, $\epsilon = 0$, $\epsilon_i = 10$, $\delta = \delta_i = 1$. We also consider  $z_i(0)$ and $\phi_i(0)$ as random parameters as in the previous figure.}
\label{figura11}
\end{figure}

An interesting process occurs when we consider the system as a symmetric TLS and the environment as asymmetric TLSs. As the left panel of Fig. \eqref{figura1} shows, an isolated symmetric TLS oscillates between the left and the right states, as it is well known. Interestingly, when the asymmetry of the environment is switched on, it affects the system by transferring its asymmetry to the central TLS (see Fig. \eqref{figura11}). Therefore, we have managed to design an
interacting model which turns
an isolated symmetric TLS into an asymmetric one.

\section{Conclusions and final remarks}\label{section6}
Along this manuscript, we have presented a Hopf-based formalism to obtain a Meyer-Miller-Stock-Thoss mapping to study the dynamics of two-level systems, turning the quantum Hamiltonian operator into a Hamiltonian function.

After showing the description of two-level systems using canonically conjugate variables, we have presented an interacting two-level system model, 
bilinearly coupling the corresponding population differences. We have shown that the dynamics behave as a Gross-Pitaevskii-like one, the non-linear term being the coupling between population differences. 
We also exposed a transition between oscillatory and tunneling-suppressed dynamics that can be observed by varying the coupling constant, $\Lambda$.

Our interacting model has also been extended to a system plus environment model, considering the environment as a collection of two-level systems through a Caldeira-Legget-like formalism. In this context, our system shows a chaotic behavior forcing us to employ averages of the population differences to obtain reliable results. Among the results, we have noted the tunneling-suppressed dynamics in the strong coupling limit and the usual damping effect similar to that of a harmonic oscillator bath in the weak coupling limit. Even more, we have managed to design an interacting model which turns an isolated symmetric two-level system into an 
environment-assisted asymmetric one.

Finally, we would like to remark that the formalism here introduced can be easily adapted to different fields in the context of open quantum systems, ranging from 
quantum computation or condensed matter, for instance. Specifically, we are currently employing the techniques here introduced to tackle the problem of the role of parity violation in a model of interacting chiral molecules. 

\section*{Acknowledgements}
Discussions with J. A. Pons are gratefully acknowledged.
D. M. -G. acknowledges Fundación Humanismo y Ciencia for financial support. D. M. -G. and P. B. acknowledge Generalitat Valenciana through PROMETEO PROJECT CIPROM/2022/13.

\bibliographystyle{unsrt}
\bibliography{referencias.bib}

\end{document}